\documentclass[twocolumn,showpacs,amsmath,amssymb,prl,superscriptaddress]{revtex4}
\usepackage{graphicx}
\usepackage{bm}
\usepackage{dcolumn}

\begin{document}
\title{ Proximity of Iron Pnictide Superconductors to a Quantum  Tricritical Point} 

\author{G. Giovannetti}
\affiliation{ISC-CNR and Dipartimento di Fisica, Universit\`a di Roma 
``La Sapienza'', P.  Aldo Moro 2, 00185 Roma, Italy.}
\author{C.  Ortix}
\affiliation{Institute for Theoretical Solid State Physics, IFW Dresden, 01171 Dresden, Germany}
\author{M. Marsman}
\affiliation{Faculty of Physics and Center for Computational Materials Science,
University Vienna, Sensengasse 8/12, A-1090, Vienna, Austria.}
\author{M. Capone}
\affiliation{ISC-CNR and Dipartimento di Fisica, Universit\`a di Roma 
``La Sapienza'', P.  Aldo Moro 2, 00185 Roma, Italy.}
\author{J. van den Brink}
\affiliation{Institute for Theoretical Solid State Physics, IFW Dresden, 01171 Dresden, Germany}
\author{J. Lorenzana}
\affiliation{ISC-CNR and Dipartimento di Fisica, Universit\`a di Roma 
``La Sapienza'', P.  Aldo Moro 2, 00185 Roma, Italy.}
\date{\today}

\begin{abstract}
We determine the nature of the magnetic quantum critical point in the
doped LaFeAsO using a set of constrained density functional
calculations that provide {\it ab initio} coefficients for a Landau order
parameter analysis. The system turns out to be remarkably close to 
a quantum tricritical point, where the nature of the phase transition changes
from first to second order. We compare with the effective field theory
and discuss the experimental consequences.
\end{abstract}
\maketitle

\begin{figure}
\includegraphics[width=1\columnwidth]{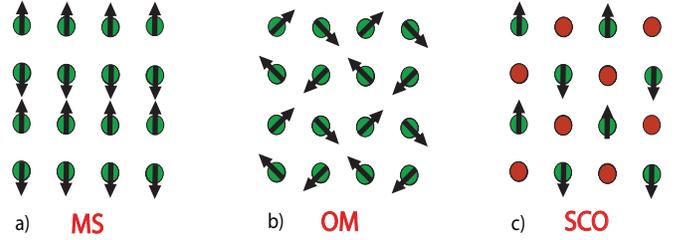}
\caption{(Color online) Competing magnetically ordered states in LaOFeAs.}
\label{fig:competingorders}
\end{figure}

Layered FeAs materials have been extensively studied since the
discovery of superconductivity with a transition temperature up to 28
K in LaFeAsO$_{1-x}$F$_{x}$ and  exceeding 50K in related
compounds. Neutron scattering experiments~\cite{cru08} have shown that
in the FeAs planes of the layered parent compound long-range magnetic
stripe order develops [Fig.~\ref{fig:competingorders}(a)]  which was predicted by density functional theory~\cite{don08}.  Upon doping the compound magnetic order is suppressed and disappears at zero
temperature at a quantum critical point (QCP). The phase diagram is strikingly similar to that of heavy fermions where the QCP is inside
or on the edge of the superconducting dome~\cite{mat98mon07}. 
The role of a QCP in promoting superconductivity has been stressed in  
heavy fermions\cite{col07} and  cuprates\cite{she01}
which renders the nature of the QCP of fundamental importance.

Empirically the magnetic {\em thermal} transitions show basically two types of behavior~\cite{kre08,jes10}. In some compounds, particularly of the 122 family, such as SrFe$_2$As$_2$, the transition is first order, while in compounds of the 1111 family such as LaFeAsO it appears second order like. This suggests that the iron pnictides are close to a tricritical point, {\em i.e.} a point in the $T$-$x$ plane, with $x$ a non-thermal parameter, where the nature of the transition changes from first to second order. 
It is in principle conceivable that by changing two non-thermal parameters the tricritical point is driven to zero temperature producing a quantum tricritical point (QTCP) where a range of unconventional quantum critical phenomena is expected to occur. Accidental proximity to such a QTCP can also dominate the finite temperature crossovers and is believed to occur in several compounds~\cite{mis09jak10}. 

Whether such a scenario is viable for the doped iron pnictides can be
determined from a Landau order parameter analysis if the coefficients
that appear in the expansion are known. Identifying all possible and
relevant magnetic phases~\cite{lor08} close to the potential
tricritical point, we have determined all the coefficients of the
Landau theory in LaFeAsO as a function of doping from first principles, computing total energies in a constrained density functional approach~\cite{ded84} within the local-density approximation (LDA)~\cite{hoh64,koh65,per81}. 
We find that LaFeAsO is surprisingly close to a QTCP which will strongly affect superconducting and normal state properties. The resulting effective field theory ~\cite{xu08} reveals that at this critical point an Ising and a continuous order parameter vanish concomitantly. Upon doping a structural transition must therefore be very close to the magnetic one, as indeed is found in the experimental phase diagram of LaFeAsO.



The Landau theory is based on the assumption that the dominant magnetic instability of  FeAs planes is at momentum $(\pi,0)$ and  $(0,\pi)$ where we use a notation with a single Fe per unit cell and take the Fe-Fe distance $a\equiv 1$. 
Besides the well known magnetic stripe phase one needs to include here also the phase with orthogonal magnetic moments, termed orthomagnetic (OM), and the phase with spin and charge order (SCO) 
(c.f. Fig.~\ref{fig:competingorders})~\cite{lor08}. In accord with the
Landau analysis, these states appear as low-lying energy phases in  
microscopic computations\cite{ere10}, {\it ab initio}~\cite{ynd09sha09}
and density matrix renormalization studies~\cite{ber10}.  


Within the Landau expansion the order parameter is given by the six Cartesian components of the Fourier transform of the magnetization at the relevant wave vectors. We need to keep the expansion up to the sixth power of the order parameter in order to account for second and weakly first order phase transitions close to a tricritical point. Using symmetries it is found that the Landau expansion has 7 independent coefficients 
which we need to determine\cite{lor08}. 
The energies of the three ordered phases shown in Fig.~\ref{fig:competingorders} 
as a function of their total magnetization, $M_T$,  completely determine the 7 Landau coefficients. The energies 
are given by 
\begin{eqnarray}
  \label{eq:ener_phases}
  \delta f_{MS}&=&\frac\alpha2 M_T^2 + B_1 M_T^4 + G_1 M_T^6\nonumber\\
  \delta f_{OM}&=&\frac\alpha2 M_T^2 + B_2 M_T^4 + G_2 M_T^6\\
  \delta f_{SCO}&=&\frac12\left(\frac\alpha2 M_T^2 + B_3 M_T^4 +G_3 M_T^6\right).
\nonumber
\end{eqnarray}
%
For the SCO state the factor $1/2$ accounts for the fact that only one half of the Fe sites are magnetized.

For the LDA computations  we used the Vienna \textit{ab-initio}
simulation package (VASP)\cite{kre96}.  The Kohn-Sham equations in the
self-consistent calculations have been solved using the projector
augmented wave method~\cite{kre99} with the valence
pseudo-wave-functions expanded in a plane wave basis set with a
cut-off energy of 500 eV. All the integrations in the Brillouin zone
are performed initially with a Gaussian smearing method and then
checked with a tetrahedron scheme\cite{blo94} using  a sampling grid  of
$10\times 10\times 6$ k-points. 

For iron pnictides the method of choice is  LDA because it provides
magnetic properties closer to experiment than, for example,
generalized gradient  approximation~\cite{maz08}. We use experimental
lattice and internal parameters fixed at zero doping~\cite{cru08} with
symmetry group P4/nmm, constructing our unit cell with 4 Fe sites to
be able to allocate the mentioned magnetic structures. We also
performed calculations changing the $z$ coordinate of the As as
explained below.  Electron doping has been introduced in our
computations by the virtual crystal approximation\cite{leb09}.

We fixed both the modulus and direction of the magnetization to the patterns dictated by the Landau theory by implementing in VASP  the possibility to perform LDA contrained calculations\cite{ded84} and computing the total energy as a function of M$_T$. The local magnetic moments, needed to implement the constraint,  where found by integration of the magnetization density in atomic Wigner-Seitz spheres centered at the Fe sites. The energy versus $M_T$ curves were fitted with expressions Eq.~(\ref{eq:ener_phases}) to determine the Landau coefficients.  The minimum of the Landau energy determines the equilibrium magnetization.  

The MS phase breaks $C_4$ symmetry thus we expect that the lattice will distort as indeed observed experimentally. However we are interested in the behavior close to the QCP between the magnetic and non-magnetic state, where the orthorhombicity becomes negligible~\cite{cru10} thus for simplicity we neglect this effect.  


\begin{figure}
\includegraphics[width=\columnwidth]{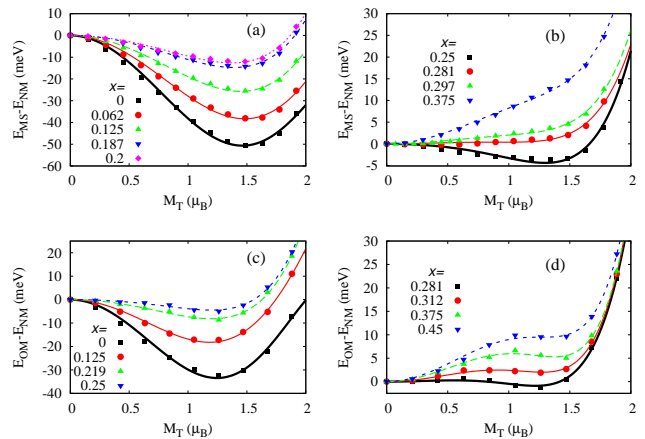}
\caption{(Color online) Energy per Fe for the MS (a),(b) and
  the OM phase (c),(d) as a
  function of the Fe magnetic moment for different doping concentration
   $x$. The points are the results of LDA calculations
  whereas the continuous lines are fits using
  Eq.~(\ref{eq:ener_phases}). } 
\label{fig:constrainedene}
\end{figure}

Fig.~\ref{fig:constrainedene} shows the total
energy per Fe for the MS and OM state as a function of the local Fe
magnetic moment $M_{T}$ and different doping concentrations. The zero of the
energy is taken at the $M_{T}=0$ state. The points are
the LDA data and the lines the Landau fits.  In general we find that
the latter provides an excellent fit to the data,  even when $M_T$ is
not small.
For the MS in the undoped case ($x=0$), the energy curve shows a clear deep
minimum for $M_{T} \sim 1.5 \mu_{B}$
[Fig.~\ref{fig:constrainedene}(a)]. By increasing the F-doping, the
minimum moves to slightly lower moment values and 
eventually disappears for $x>0.3$ 
[Fig.~\ref{fig:constrainedene}(b)].  
The energy is extremely flat close to the critical point with the
large moment state almost degenerate with the zero magnetic state
and without the appearance of a noticeable  barrier. 
This behavior already indicates proximity to a QTCP and an anomalously
``soft'' magnetism at the transition.
For the OM state Fig.~\ref{fig:constrainedene}(c),(d)
the behavior 
has a more pronounced first order character. The
metastable minimum persists up to large dopings with a sizeable
barrier separating the large magnetization state from the low
magnetization state. Finally for the SCO state (not shown)  the
evolution is that of a typical second order phase transition.  

From the statistical mechanics point of view the LDA approximation
is a mean-field theory. Thus in the second order region of the phase
diagram, at the LDA level, one finds classical critical
exponents. The order parameter as a function of a non-thermal
parameter should vanish as $|x-x_\alpha|^\beta$ with $\beta_{LDA}=1/2$ far from
the tricritical point in the second order region and
$\beta_{LDA}=1/4$ at the tricritical point. 
In the critical region the energy landscape is very flat which is a big obstacle for the converge of conventional LDA computations. We avoid such complications by constrained computations that allow to extract the intrinsic LDA behavior.

\begin{figure}
\includegraphics[width=1\columnwidth]{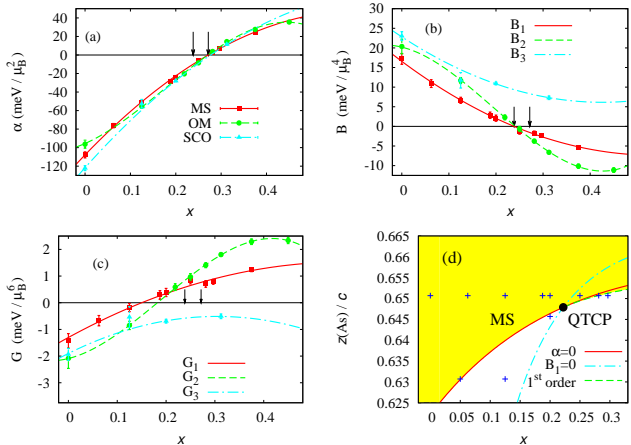}
\caption{(Color online)(a),(b),(c): 
Landau parameters for the three possible magnetic phases as function
of the doping concentration $x$. Dots are the numerical results while
the lines are fits as explained in the text. The left (right) arrow indicate
the point where $B_1=B_2=0$ ($\alpha=0$). 
Panel (d) shows the LDA phase diagram for the MS in the $z$(As)-doping plane.}
\label{fig:Landauparameters}
\end{figure}

The susceptibility $\alpha^{-1}$ of the nonmagnetic phase at momentum $(\pi,0)$ or $(0,\pi)$ should be independent of the phase under consideration, thus a single parameter $\alpha$ appears in Eqs.~(\ref{eq:ener_phases}). As a consistency check we allowed for different values of $\alpha$ in the fits and found that indeed $\alpha$ converges to practically the same values as a function of doping except for $x\sim 0$ and 0.5 where higher order terms in the expansion become important [Fig.~\ref{fig:Landauparameters}(a)]. The vanishing of $\alpha$ at a critical doping $x_\alpha$ determines the limit of stability of the paramagnet coming from large $x$. For the experimental lattice
constants we find $\alpha \sim 0.26 (x-x_\alpha)$ eV/$\mu_B^2$ with  $x_\alpha=0.27$.

\begin{figure} 
\hspace{0.25cm}
\includegraphics[width=1\columnwidth]{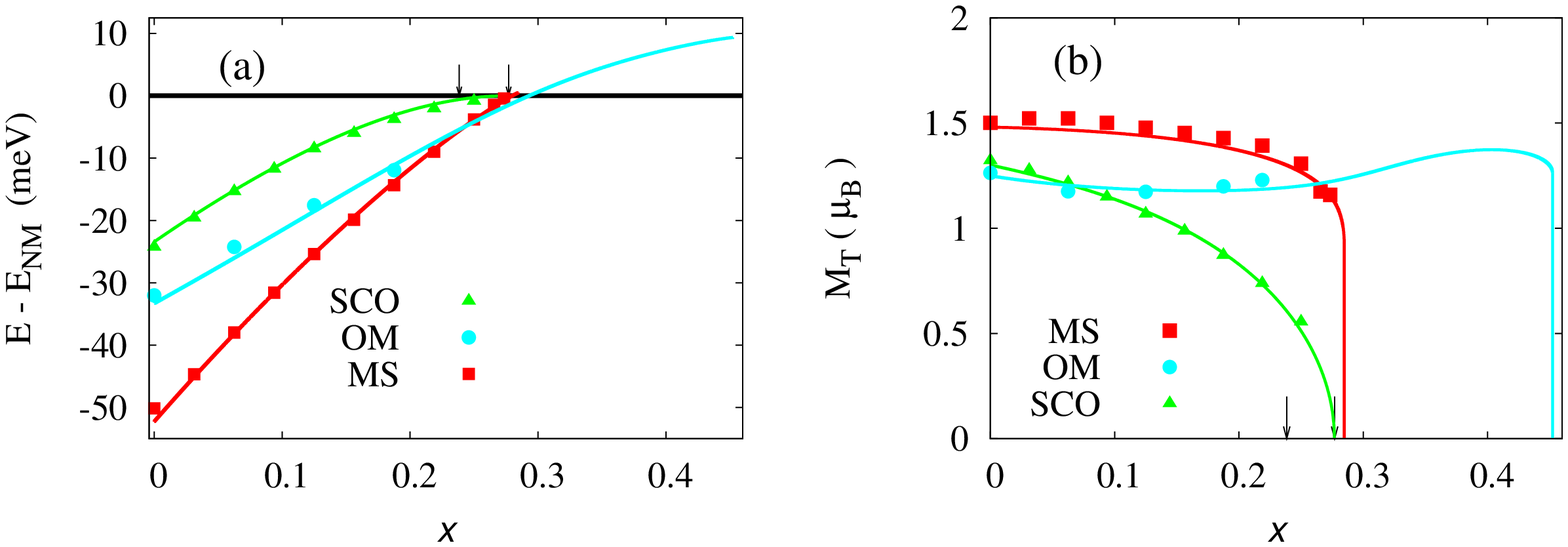}
\caption{(Color online) Energy per Fe (a) and $M_T$ 
  (b) in the  unconstrained  LDA (dots) and the
  Landau theory (lines). The meaning of arrows is the same as in
  Fig.~\ref{fig:Landauparameters}.}   
\label{fig:energy}
\end{figure}

Fig.~\ref{fig:Landauparameters}(b),(c) show the behavior of the quartic and sixth order Landau parameters as a function of the doping concentration $x$. The behavior is rather smooth and can be simply captured assuming a quadratic polynomial expansion except for the OM where a cubic term becomes important far from the critical point.  
For the MS and the OM phases we find that the  quartic coefficient $B$ of the Landau expansion becomes negative above $x_B=0.24<x_\alpha$. Therefore the transitions from the non-magnetic state to the MS and OM phase are first order while for the SCO state $B_3>0$ and the transition is a conventional second order one.  
One can judge the relevance of the QTCP in the thermal crossovers by computing the  height of the energy barrier  at the point in which the
magnetic and non-magnetic solution become degenerate.  For the MS the barrier from the fits is nominally $\sim 2$K per Fe atom which is much below the limit of accuracy of the computation, thus for all practical propose the MS-non-magnetic transition occurs at a QTCP in LDA. 
The low barrier reflects an almost vanishing metastability range around the transition and the physics will be dominated by QTCP behavior. 
The OM state has a larger range of metastability (c.f. Fig. \ref{fig:energy}) but still with a negligible barrier at the transition point ( $ \sim 10 K$). 

Fig.~\ref{fig:energy}(a) shows the energy of the different phases. The
end of the line indicates the spinodal point, {\em i.e.} the point at
which the state disappears as a saddle point solution of the Landau
equations. The MS stripe is the most stable phase except close to the
transition to the paramagnet where the OM phase becomes stable in a
small doping interval. The OM preserves
$C_4$ symmetry and has the same structure factor as an incoherent
superposition of $(\pi,0)$ and $(0,\pi)$ twins of 
the MS state, so it is difficult to
distinguish with magnetic neutron scattering alone. It is interesting
that for 122 compounds a state with magnetic order but without
detectable orthorhombicity has been reported~\cite{par09} which can be
taken as a signature of the OM
state. However this is the region where the orthorhombicity is expected to be smaller and further experimental work is required to clarify this point. 

Fig.~\ref{fig:energy}(b) shows $M_T$ vs. doping. Despite the abrupt suppression of the order parameter from $M_{T} \sim 1 \, \mu_{B}$ to $M_{T} \equiv 0$ for the MS, the transition is weakly first order. The OM magnetization shows a sharper first order behavior although, as discussed above, the barrier is very small.  The SCO behaves as a typical second order transition. The behavior of the order parameter of the MS is consistent with  the sudden drop of the magnetization as a function of doping observed in this compound~\cite{lue09}. Such behavior reinforces our conclusion that the system is close to a QTCP although our critical doping is overestimated as discussed below. 
The fact that the OM state lies only less than 10meV above the paramagnetic state for a large range of doping suggests that fluctuations to this state may be the most relevant ones in the superconducting region. 

 
The LDA critical doping $\sim 0.26 $ is larger than the experimental one 0.05 $\sim$ 0.06 in this compound. Since LDA neglects fluctuations the Landau parameters derived should be considered to be ``bare'' parameters. One well known effect of fluctuations is to reduce the stability of the ordered phases~\cite{sac01,and01}shifting the critical doping in the correct direction. Also  because of the small energies involved, the actual transition will be sensitive to details as the choice of the functional or changes due to the relaxation of the structure. In particular it is
 known that the magnetism is very sensitive to $z$(As) the $z$ coordinate of the As in the unit cell~\cite{yin08,maz08,ynd09sha09}. Relaxing $z$(As) results in a decrease of around $0.1 c$ ($c=0.87\AA$). We made also computations with selected values of $z$ and constructed the zero temperature phase diagram shown in Fig.~\ref{fig:Landauparameters}(d) which shows the location of the QTCP in the $z$(As)-doping plane.  We see that a decrease of $z$(As) indeed shifts the transition to the right range of doping moving from the weakly first order region (right of the QTCP) to the second order region. We also show the line $B_1=0$. We see that even if one takes a relaxed $z$ and a smaller critical doping still the transition is dominated by the proximity to  this line. Indeed we find that the $B_1$ coefficient depends weakly on $z(As)/c$ so for doping $x=0.05$ and $z/c=0.6307$ we find $B_1\sim 10$ meV$/\mu_B^4$. Taking a characteristic  moment of $M^*\sim 0.5\mu_B$ this implies a negligible energy scale $B_1 (M^*)^4 \sim 7$K which again points to soft magnetism due to proximity to the QTCP.  

Ref.~\cite{xu08} has analyzed the field theory describing the quantum phase transition in iron pnictides in terms of fields ${\vec \phi}_1$ and
${\vec \phi}_2$ representing N\'eel order parameters in different sublattices and related to our order parameter by  ${\vec \phi}_{1,2}={\bf M}_1 \pm {\bf M}_2$.  
The second and fourth order terms are precisely equivalent to ours: $\sum_{i=1}^2 u|{\vec \phi}_i|^4 +\gamma_1 |{\vec \phi}_1|^2|{\vec \phi}_2|^2 - \alpha_X ({\vec \phi}_1.{\vec \phi}_2)^2$. In terms of our coefficients $u=B_3 / 2 $, $\gamma_1= B_2-B_3$ and $\alpha_X=B_2-B_1$. Increasing the doping, for the experimental atomic positions, $\alpha_X$ changes from positive to negative at $x_B$ [Fig.~\ref{fig:Landauparameters}(b)] immediately before the transition. This is consistent with the appearance of the OM state in a small doping interval as shown above. A small $\alpha_X$ implies a structural transition very close to the magnetic transition as a function of doping~\cite{xu08} which is consistent with the experimental phase diagram of doped LaOFeAs~\cite{lue09}. 


To conclude, we have shown that magnetism in an iron-pnictide superconductor is surprisingly close to a QTCP and we have determined {\it ab initio} the coefficients of a Landau expansion around it. The energy landscape is anomalously flat close to the zero temperature magnetic-non-magnetic transition giving rise to a very soft behavior of the order parameter in the sense that can experience large changes as a result of weak perturbations. We believe frustration plays and important role in this result as it tends to turn Stoner like transitions into weakly first order ones~\cite{lor08}. 
While we have focussed on a 1111 compound one expects similar behavior to occur in 122 compounds. In particular it has been shown that magnetism can be quenched and superconductivity appears in the undoped compound by applying pressure~\cite{ali09}. Thus experiments exactly at the QTCP should be possible by a combination of pressure and doping to elucidate the effect of this unusual critical behavior on
superconducting and normal state properties. The degeneracy of the ground state close to a QCP is believed to boost superconductivity as a way to remove the residual entropy\cite{col07,cap04}. This effect should be enhanced close to a QTCP, where the degeneracy is even larger, and may play an important role in determining the high critical temperature of  iron pnictides.

\acknowledgments
GG thanks A. Stroppa for useful discussions.
This work is supported by IIT-Seed project NEWDFESCM and by CINECA
who allocated computer time. M.C. and G.G. are financed by ERC through the Starting Independent Grant ``SUPERBAD'', Grant Agreement No. 240524


\begin{thebibliography}{10}

\bibitem{cru08}
C. de~la Cruz {\it et~al.}, Nature {\bf 453},  899  (2008).

\bibitem{don08}
J. Dong {\it et~al.}, Europhys. Lett. {\bf 83},  27006  (2008).

\bibitem{mat98mon07}
N.~D. Mathur {\it et~al.}, Nature {\bf 394},  39  (1998).
P. Monthoux, D. Pines, and G.~G. Lonzarich, Nature {\bf 450},  1177  (2007).

\bibitem{col07}
P. Coleman in {\it Handbook of Magnetism and Advanced Magnetic
  Materials, Vol 1}, edited by H. Kronmuller and S. Parkin (Wiley
and Sons, New York, 2007). 

\bibitem{she01}J. H. She and J. Zaanen, Phys. Rev. B, {\bf 80}, 184518
  (2009) and references therein. 

\bibitem{kre08}
C. Krellner {\it et~al.}, Phys. Rev. B {\bf 78},  100504  (2008).

\bibitem{jes10}
A. Jesche {\it et~al.}, Phys. Rev. B {\bf 81},  134525  (2010).

\bibitem{mis09jak10}
T. Misawa, Y. Yamaji, and M. Imada, J. Phys. Soc. of Japan
  {\bf 78},  084707  (2009).
P. Jakubczyk, J. Bauer, and W. Metzner, Phys. Rev. B {\bf 82},  045103  (2010).

\bibitem{lor08}
J. Lorenzana, G. Seibold, C. Ortix, and M. Grilli, Phys. Rev. Lett. {\bf 101},
  186402  (2008).

\bibitem{ded84}
P.~H. Dederichs, S. Bl\"ugel, R. Zeller, and H. Akai, Phys. Rev. Lett. {\bf
  53},  2512  (1984).

\bibitem{hoh64}
P. Hohenberg and W. Kohn, Phys. Rev. {\bf 136},  B864  (1964).

\bibitem{koh65}
W. Kohn and L.~J. Sham, Phys. Rev. {\bf 140},  A1133  (1965).

\bibitem{per81}
J.~P. Perdew and A. Zunger, Phys. Rev. B {\bf 23},  5048  (1981).

\bibitem{xu08}
C. Xu, M. M\"uller, and S. Sachdev, Phys. Rev. B {\bf 78},  020501  (2008).

\bibitem{ere10}
I. Eremin and A.~V. Chubukov, Phys. Rev. B {\bf 81},  024511  (2010).

\bibitem{ynd09sha09}
F. Yndurain and J.~M. Soler, Phys. Rev. B {\bf 79},  134506  (2009).
S. Sharma {\it et~al.}, Phys. Rev. B {\bf 80},  184502  (2009).

\bibitem{ber10}
E. Berg, S.~A. Kivelson, and D.~J. Scalapino, Phys. Rev. B {\bf 81},  172504
  (2010).

\bibitem{kre96}
G. Kresse and J. Furthm\"uller, Phys. Rev. B {\bf 54},  11169  (1996).

\bibitem{kre99}
G. Kresse and D. Joubert, Phys. Rev. B {\bf 59},  1758  (1999).

\bibitem{blo94}
P.~E. Bl\"ochl, Phys. Rev. B {\bf 50},  17953  (1994).

\bibitem{maz08}
I.~I. Mazin {\it et~al.}, Phys. Rev. B {\bf 78},  085104  (2008).

\bibitem{leb09}
S. Leb\`egue {\it et al.}, New J. Phys. {\bf 11} 025004 (2009).

\bibitem{cru10}
C. de~la Cruz {\it et~al.}, Phys. Rev. Lett. {\bf 104},  017204  (2010).

\bibitem{par09}
J.~T. Park {\it et~al.}, Phys. Rev. Lett. {\bf 102},  117006  (2009).

\bibitem{lue09}
H. Luetkens {\it et~al.}, Nature Materials {\bf 8},  305  (2009).

\bibitem{sac01}
S. Sachdev, {\it Quantum phase transitions} (Cambridge University
Press, Cambridge, 2001).

\bibitem{and01}
S. Andergassen {\it et al.}, Phys. Rev. Lett. {\bf
  87},  056401  (2001).

\bibitem{yin08}
Z.~P. Yin {\it et~al.}, Phys. Rev. Lett. {\bf 101},  047001  (2008).

\bibitem{ali09}
P.~L. Alireza {\it et~al.}, J. Phys.: Condens. Matter {\bf 21},
  012208  (2009).

\bibitem{cap04}
M. Capone {\it et~al.}, Phys. Rev. Lett. {\bf 93}, 047001 (2004).
\end{thebibliography}

\end{document}